\documentstyle[epsfig]{aipproc}

\title {A Gamma-Ray Burst Bibliography, 1973-2001}
\author{K. Hurley}
\address{UC Berkeley\\ Space Sciences Laboratory\\ Berkeley, CA 94720-7450}

\begin{document}

\maketitle

\begin{abstract}

On the average, 1.5 new publications on cosmic gamma-ray bursts
enter the literature every day.  The total number now exceeds 5300.
I describe here a relatively complete bibliography which is on the web, and 
which can be made available
electronically in various formats.
\end{abstract}

\section{Introduction}

I have been tracking the gamma-ray burst literature for about the past twenty-one years, keeping the authors, titles, references, and key subject words in a machine-readable file.  The present version updates previous ones reported in 1993 \cite{hurley1}, 1995 \cite{hurley2},1997 \cite{hurley3}and 1999 \cite{hurley4}. In its current form, this information is in a Microsoft Word 97 "doc" format.  My purpose in doing this was first, to be able to retrieve rapidly any articles on a given topic, and second, to be able to cut and paste references into manuscripts in preparation.  The following journals have been scanned on a more or less regular basis starting with the 1973 issues:\\
\linebreak
Advances in Physics\\
Annals of Physics\\
Astronomical Journal\\
Astronomische Nachrichten\\
Astronomy and Astrophysics (including Supplement Series)\\
Astronomy and Astrophysics Review\\
Astronomy Letters (formerly Soviet Astronomy Letters)\\
Astronomy Reports (formerly Soviet Astronomy)\\
Astrophysical Journal (letters, main journal, and supplements)\\
Astrophysical Letters and Communications\\
Astrophysics and Space Science\\
ESA Bulletin\\
ESA Journal\\
IAU Circulars\\
IEEE Transactions on Nuclear Science\\
Journal of Astrophysics and Astronomy\\
Monthly Notices of the Royal Astronomical Society\\
Nature\\
Nuclear Instruments and Methods in Physics Research Section A\\
Observatory\\
Physical Review (main journal A and letters)\\
Proceedings of the Astronomical Society of Australia\\
Publications of the Astronomical Society of Japan\\
Publications of the Astronomical Society of the Pacific\\
Reports on Progress in Physics\\
Science\\
Scientific American\\
Sky \& Telescope\\

In addition, the following journals either have been scanned, but less regularly in the past, or
in some cases, are no longer being scanned:\\
\linebreak
Annals of Geophysics\\
Astrofizika\\
Astroparticle Physics\\
Bulletin of the American Astronomical Society\\
Bulletin of the American Physical Society\\
Bulletin of the Astronomical Society of India\\
Chinese Astronomy and Astrophysics\\
Chinese Physics Letters\\
Cosmic Research\\
Journal of Atmospheric and Terrestrial Physics\\
Journal of the British Interplanetary Society\\
Journal of the Royal Astronomical Society of Canada\\
New Astronomy\\
Progress in Theoretical Physics\\
Solar Physics\\
Soviet Physics\\

The above lists are not exhaustive.  For example, where theses or internal reports have come to my attention, I have included them, too.  To be included, an article had to have something to do with GRB or SGR  theory, observation, or instrumentation, or be closely related to one of these topics (e.g., merging neutron stars, AXPs, high-z supernovae, etc.), and must have been published.  With only a few exceptions, preprints or internal reports which were never published have not been included.

\section{Organization of the Bibliography}

The overall organization is chronological by year.  Within a given year, articles published in journals are listed first, in alphabetical order by first author.  Then come theses and conference proceedings articles.  The latter are listed in the order in which they appear in the proceedings.  The entries are numbered consecutively, so that paper copies which are kept on file can be retrieved quickly.  However, to avoid having to renumber this entire file when a new article is added, numbers are skipped at the end of each year and reserved for later inclusion.  The complete author list follows, as it appears in the journal, along with the title, journal, volume number, page number, and year.  A line containing key words follows this.  These are generally not the same key words as the ones listed in the journal, nor are they taken from the title or any particular list.  Rather, they are meant to reflect the true content of the article, and provide a list of machine-searchable topics.  In general, however, key words have not been included for conference proceedings articles.  An example of an entry is the following:\\
\linebreak
5163.	Guetta, D., Spada, M., and Waxman, E., On the Neutrino Flux from Gamma-Ray Bursts, Ap. J. 559, 101, 2001\\
Key Words: p-gamma interactions, photomeson production, 10\^14 eV neutrinos

\section{A Few Interesting Statistics}

The number of articles published each year since 1973 is shown in figure 1.  Starting with one article per month  in 1973, it began to exceed one per day in 1994, and reached over 1.5 per day in 2000, enough, in principle, to base an entire journal on.
Several milestones are indicated as the probable causes of sudden increases in the number of publications per year.  Note that there are still about as many papers published as there are gamma-ray bursts observed.  The cumulative total is shown in Figure 2.  The cutoff date is mid-2001.  At any given time, there
may be about 100 articles waiting to be entered into the file, so the completeness, including an estimate of the number of articles which were missed
for any reason, is about 98\%.

\begin{figure}
\psfig{file=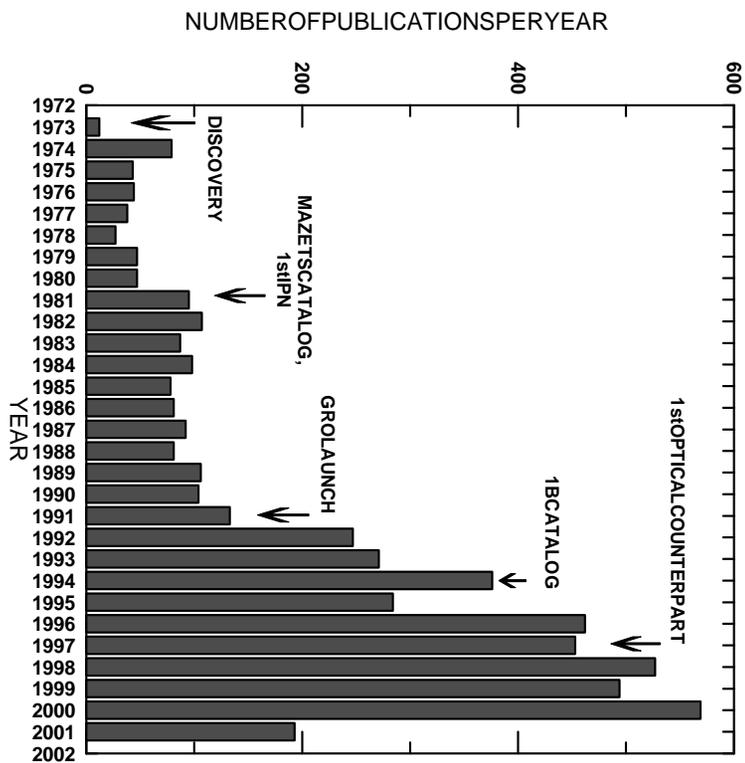, width=10cm}
\caption{The number of publications by year.}
\label{fig1}
\end{figure}

\begin{figure}
\psfig{file=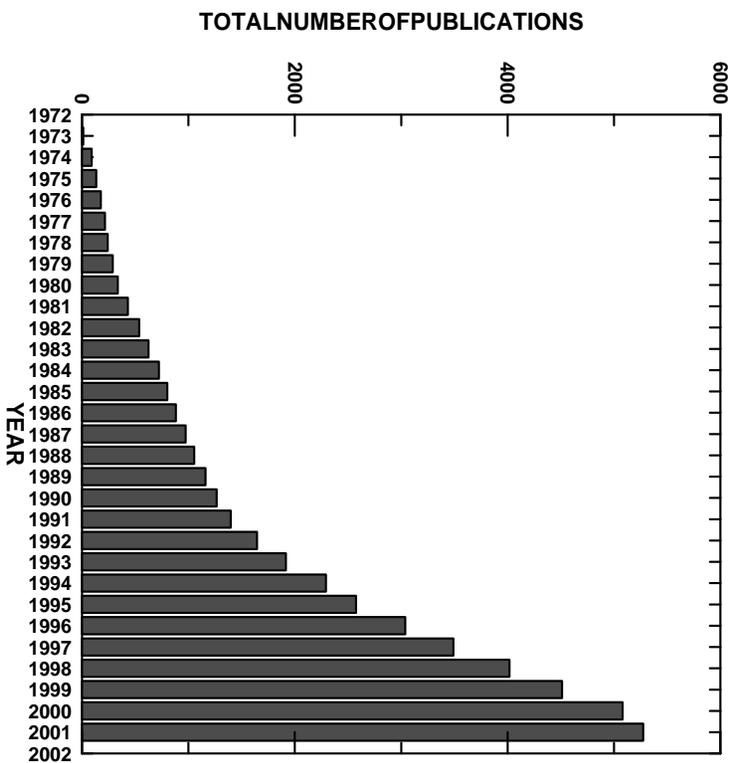, width=10cm}
\caption{The cumulative number of publications by year. }
\label{fig1}
\end{figure}

The volume of the literature (it would take about 600 pages simply to print out  the bibliography)
has necessitated the development of a program which can search for and extract particular titles.  I have written such a program in Microsoft Word Basic (a variant of the BASIC programming language).  It allows one to extract all titles between two dates whose entries contain a particular key phrase,
key word, or author, and write them to a separate file.  

\section{Availability}

A web version of this bibliography may be found at ssl.berkeley.edu/ipn3/index.html.  However, although the bibliography
is updated on an approximately daily basis, the most up-to-date version is usually not at the website. It is available in plain ASCII, "doc", and "rich text format" (rtf) format files, which can be sent to anyone interested, as can the Word Basic program.  Please contact me at khurley@sunspot.ssl.berkeley.edu to request copies, and indicate your preference for the format.  I would appreciate it if users would communicate errors and omissions to me.

This work was carried out under JPL Contract 958056.

\end{document}